\documentclass[ twocolumn ,secnumarabic, amssymb, amsmath, aps, prl]{revtex4}
\usepackage{graphicx}
\usepackage{amssymb}
\usepackage{epstopdf}
\usepackage{epsfig}
\def \be{\begin{displaymath}}
\def \ee{\end{displaymath}}              % equation without number

\def \ben{\begin{equation} }
\def \een{\end{equation}   }            % equation with number

\def \bea{\begin{eqnarray*}}             % equation array without numbers
\def \eea{\end{eqnarray*}}

\def \bean{\begin{eqnarray}}             % equation array with numbers
\def \eean{\end{eqnarray}}

\def \Ref#1{(\ref{#1})}

\def \j{ {\mathbf j} }
\def \E{ {\mathbf E} }

\def \grad{ \nabla }

\def \Gap{ {\tilde G} }

\def \fr#1#2{ \frac{#1}{#2} }

\begin{document}

\title{Residual resistance in two-dimensional, microwave driven Hall systems}

\author{Rochus Klesse and Florian Merz}

\address{Institut f\"ur Theoretische Physik, Universit\"at zu K\"oln, D-50937 K\"oln,
Germany}
\begin{abstract}
We address the origin of the residual resistance
observed in microwave irradiated two-dimensional electron
gases in a weak magnetic field. We study charge modulations arising
from negative photo-conductivity and show that dissipative currents 
are exponentially suppressed. Relating the exponent to the temperature
dependence of microscopic parameters taken from experiment,
we find pseudo activated behaviour. In order to obtain these results
it is essential to take into account the finite range of the Coulomb
interaction.  
\end{abstract}

\maketitle
Recently Mani {\em et al.}~\cite{mani} and Zudov {\em et
al.}~\cite{zudov} (see also Ref.~\cite{yang}) have reported that in
ultra-clean Hall systems   
the dissipative resistance is driven to exponentially small values
by sufficiently strong irradiation with microwaves. Theories addressing
these unexpected ``zero-resistance'' states are numerous and cover a
wide range of ideas \cite{durst,mirlin,ryzhii,andreev,bergeret,rest}.
A calculation by Durst {\em et al.}~\cite{durst} (see also
Refs.~\cite{mirlin,ryzhii}) revealed that the photo-conductivity 
oscillates with photon frequency and that it takes negative values in
regions  where, experimentally, the resistance is zero.  
The negative photo-conductivity leads to instability towards an
inhomogeneous, current carrying state \cite{andreev,bergeret}. Given
such a state the system can sustain a net measurement current without
voltage drop \cite{andreev}, a result which, together with
Ref.\cite{durst}, provides an explanation for the  regions of zero resistance.  

Despite this progress one of the intriguing
features remains unclear. In the ``zero-resistance'' regime an
exponentially small residual resistance remains with a temperature
dependence, which appears activated like, $R \propto \exp(-T_0/ T)$.
The measured energies $k_B T_0$ are surprisingly large. They exceed
the obvious energy scales, the cyclotron and microwave energies, by
an order of magnitude. 

In this letter we particularly address the problem of the residual
resistance and propose an alternative mechanism to thermal activation. 
Building mainly on the idea of negative local conductivity and the
resulting instability, we develop a non-local theory of  dissipative
conduction. We apply methods similar to the ones which have been used
for studying the Gunn effect \cite{volkov_kogan}. Within this
framework we investigate the global spatial structure of the electric
potential inside the system for given boundary conditions. 
We show that dissipative currents are suppressed by an exponential
factor $\exp(-\lambda/b)$, where $\lambda/2$ is the domain size in the
charge modulations and $b$ is the domain wall width. This is our main 
result. We argue that the residual resistance is suppressed by the
same factor. 
Relating $b$ to microscopic parameters measured in experiments and
taking their observed temperature dependence into account, we
find pseudo activated behaviour.
In order to obtain these results it is essential to use a finite range
interaction.

Our starting point is a phenomenological ansatz for the current
density, 
\ben\label{j_relation}  
\j = (\sigma_0 + \sigma_p(E) + \hat\sigma_H)\E - e D \grad \delta n\:,
\een 
which contains the electric drift current $\sigma_0\E$, the
Hall current $\hat \sigma_H \E$, and a microwave assisted photo-current
$\sigma_p(E) \E$ with $\E=-\grad\phi$ the local electric field, and
$E= |\E|$. $D$ is 
the diffusion constant, $\delta n$ is the deviation in the
electron density from the neutral equilibrium density, and $e$ is the
electric charge. The unusual transport behavior is captured 
in the non-linear photo-conductivity
$\sigma_p(E)$: in the zero-resistance regime and at vanishing electric  
field $\sigma_p(E)$ is large and negative~\cite{durst}.
The exact behavior of $\sigma_p(E)$ depends on many details. However,
it is clear that, for large electric fields, the photon-current
saturates and $\sigma_p(\infty)$ vanishes. Therefore we  envisage a
photo-conductivity that takes a negative minimum value at $E=0$ and
smoothly approaches zero for $E\to\infty$. A similar form has been
assumed earlier\cite{bergeret}.  

Notice that in addition to the currents driven by $\E$ there is also a diffusion
current $-e D \grad \delta n$ driven by the gradient in $\delta n$. 
It has to be taken into account, since microwave  
irradiation takes the system out of equilibrium and leads to
inhomogeneous densities. 
By writing Eq.~\Ref{j_relation} in this way we have in mind that 
there is still {\em local} equilibrium; i.e.~conductivities and diffusion
coefficient are well-defined. Moreover, in the present case 
they are, to a good approximation, independent of Fermi energy and
temperature~\cite{mani,zudov}.  
In this approximation, the relation \Ref{j_relation} together with the
continuity equation and Poisson's equation for the charge density 
$\rho = e \delta n$ 
provide a closed set of non-linear differential
equations that determine the system states.

\begin{figure}
%\vspace{.2in}
\centerline {
\epsfig{file=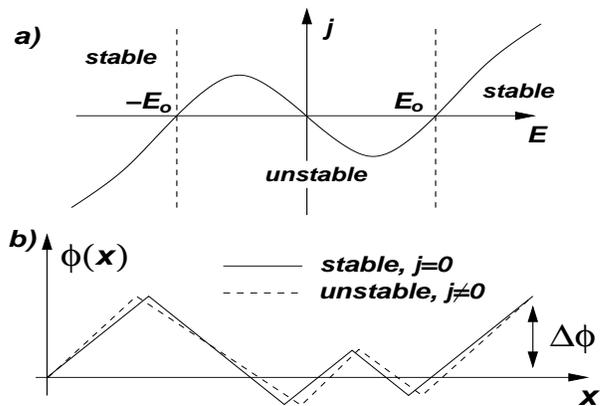,width=3.2in, height=2.2in}
}
%\vspace{.2in}
\caption{ 
a) The {\em dissipative} current $j$ as function of the electric field
for {\em homogeneous stationary states} according to Eq.~\Ref{jofe}. 
For a given current $j$ only one stable $E$ exists, except
for $j=0$, where one finds two (marginally) stable fields $\pm E_0$.
b) Stable (solid) and unstable (dashed) potential profiles for a short
range interaction. The slopes of the stable solution equal plus or
minus the critical field $E_0$. The dissipative  current $j$ of the
stable solution vanishes despite the finite potential $\Delta \phi$
drop along the system.  
}  
\label{fig-global}
\label{fig-jofe}
\end{figure}

We begin by deriving a single integral equation for the electric field
in stationary states. To keep things manageable we restrict ourself
to one-dimensional (1D) states, where the potentials 
and densities depend only on one spatial coordinate, which we take to
be $x$ throughout. 
Then, by construction, there is no $y$-component of $\E$ and hence no
Hall current in the $x$-direction. In addition, the continuity
equation requires the $x$-component of the current to be constant in
space, $j_x=const.=j$.
Next, we integrate Poisson's equation in the form
\ben\label{1d_phi}
\phi(x) = \int G(x-x') \rho(x') \:d x'\:,
\een
where the reduced 1D Green function $G(x)$ is the 
electric potential of a 1D unit line-charge $\rho(x,y) =
\delta(x)$. We take the derivative of this equation with respect to $x$,
integrate by parts and use Eq.~$\Ref{j_relation}$
to obtain the non-linear integral equation \cite{boundary_comment}
\ben\label{E_integral1D}
E(x) = -\fr{1}{D} \int G(x-x') \left[   
  (\sigma_0 + \sigma_p(E)) E  - j \right]dx'\:.
\een
The simplest solutions of this equation are states with constant electric
field $E$ throughout the system. 
In this case the dissipative current is
\ben\label{jofe}
j = [ \fr{D}{G_0} + \sigma_0 + \sigma_p(E) ] E\:,
\een
where $G_0 = \int G(x) dx$. This global $j-E$ relation is sketched
in  Fig.~\ref{fig-jofe}a. We would like to stress that this is \textit{not} a local
current-field relation but strictly holds for uniform electric fields only.
It follows from a straightforward two-dimensional (2D)  stability analysis that
homogeneous states are stable if and only if $|E|$ is larger than the 
critical field $E_0$ defined by $ D/G_0 + \sigma_0+ \sigma_p(E_0) =0 $
as indicated in Fig.~\ref{fig-jofe}. 
The associated Hall current density $\sigma_H E_0$ in
$y$-direction corresponds to the critical current density $j_0$ in
Ref.~\cite{andreev}. 

In the limiting case of a short-ranged Green function,
$G(x)=G_0\delta(x)$, Eq.~\Ref{E_integral1D} becomes local, which
allows the construction of piecewise constant solutions as depicted in
Fig.~\ref{fig-global}b. This situation here coincides with the current domain 
picture in Ref.~\cite{andreev}. Its foundation on a current-field
relation like Eq.~\Ref{jofe} has been pointed out by Begeret et
al.~\cite{bergeret}.    
Note that only states of vanishing dissipative
current $j$ and $E=\pm E_0$ can be globally stable
(cf. Fig.~\ref{fig-jofe}). Apart from that, size and configuration
of domains of critical field $\pm E_0$ are largely undetermined.

The arbitrariness of solutions is an artifact of the zero-range
interaction. Moreover, in this limit the discussion of essential properties of the
dissipative currents is not possible, as will become
clear below. Apart from this, it is questionable whether the
assumption of a short range interaction  
applies here at all. In fact, the absence of metallic
gates~\cite{mani_pc} in the experiments suggests that the screening length is large. 
We therefore set out to construct global states without having to rely on the
assumption of local interactions. We will also consider
quasi-stationary states that are weakly time-dependent. As above, we
restrict ourself to 1D states.   

Using the Green function $G_c$ of the Coulomb potential in
Eq.~\Ref{E_integral1D} makes the 
problem of finding exact solutions intractable. However, the model Green function
\ben\label{model}
\Gap(x) =  -W |x| \:,
\een
which is the exact Coulomb propagator of line charges in
two-dimensional electro-dynamics, allows a surprisingly simple construction of
non-trivial global solutions\cite{volkov_kogan}.
$\Gap$ is a good approximation to $G_c$ if one adjusts $W$ such that 
both yield the same force on the relevant length-scale. 
The improvement of choosing $\Gap$, compared to the
delta-function approximation, is the finite interaction range. 

Taking the second derivative with respect to $x$ of Eq.~\Ref{1d_phi},
with the original Green function replaced by $\Gap$, reveals that 
\ben\label{mod_poss}
-\partial^2_x \phi = \partial_x E =  2 W \rho(x)\:.
\een
Next, we take the time derivative of this equation, eliminate $\partial_t
\rho$ using the continuity equation, integrate over $x$, 
and use Eq.~\Ref{j_relation} to obtain a differential equation 
\ben\label{dyn_equation}
\partial_x^2 E - \fr{1}{D} \partial_t E  = \fr{2W}{D}\left[ (\sigma_0
  + \sigma_p(E)) E -j_0(t) \right] \:
\een
for $E(x,t)$, the $x$ component of the local, time-dependent electric field.
The integration constant $j_0(t)$ has the dimension of a
current density. Indeed, for stationary states, $j_0$ is the constant
$x$-component of the current density, as can be verified with help of
Eq.~\Ref{j_relation}. 
Actually, in the present context an equation very similar to Eq.~\Ref{mod_poss} 
has recently been discussed by Bergeret et al. \cite{bergeret}.

Stationary solutions obey 
\ben\label{motion}
\partial_x^2 E = f(E) - J\:, 
\een
where $f(E) = 2W (2\sigma_0 + \sigma_p(E)) E/D$, and $J = 2 W j_0/D$.
If $x$ is interpreted as time, this is the equation of motion of a
particle with coordinate $E$ in a force field $f(E) -J$
(cf. \cite{volkov_kogan}).  

\begin{figure}
\centerline {
\epsfig{file=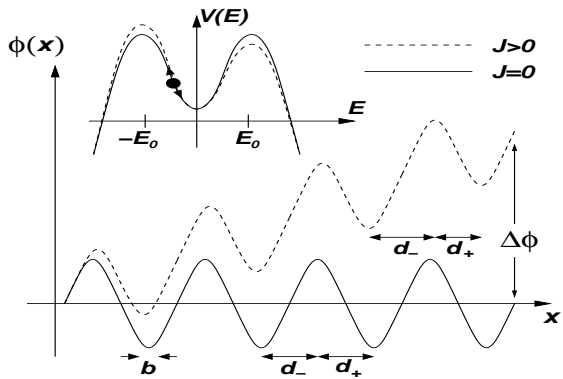,width=3.0in, height=2.0in} }
\caption{ 
Oscillations of a mechanical particle in the potential $V(E)$ 
translate into periodic modulations of the electric potential $\phi$  in the
Hall system. For vanishing dissipative current, $J=0$, the profile is ``flat'', 
while for $J>0$ (asymmetric oscillations) the potential ascends. 
}  
\label{fig-formation}
\end{figure}

For a negative zero-field conductivity $-u \equiv \sigma_0 +
\sigma_p(0)$ and sufficiently small current $J$ the  potential
$V(E)$ associated with the force in Eq.~\Ref{motion} has the shape of
an inverted double-well (Fig.~\ref{fig-formation}). 
Oscillatory orbits inside the well that almost reach the potential
maxima at $\pm E_0$ correspond to periodic field and
potential profiles.
The main differences to the solutions discussed before are the 
rounded edges and well defined domain lengths $d_-$ and $d_+$. 
The rounding width $b$ can be estimated by the frequency $\Omega$ of
harmonic oscillations near the minimum of the potential, $b \approx \pi / \Omega$.
The domain lengths $d_-$ and $d_+$ are given by the dwelling time of the
particle near the left or right maximum. They can therefore be
estimated by $d_\pm \approx 2/\omega \ln E_0/q_\pm$, where $q_\pm$ is
the minimum distance of the particle orbit to $\pm E_0$, and $\omega$
the frequency of oscillations in the inverted potential near $\pm E_0$.
In parabolic approximation of $\sigma_p(E)$ the frequencies $\omega$
and $\Omega$ differ only by a factor $\sqrt{2}$, s.t. the estimates
for $b$ and $d_\pm$ lead to the simple relation
\ben\label{exponential}
\fr{d_\pm}{b} \sim  \ln \fr{E_0}{q_\pm}\:.
\een
  
For vanishing current, $V(E)$ is symmetric in $E$
and $d_+ = d_-$. This is the ``flat'' solution shown in
Fig.~\ref{fig-formation}. A small but finite dissipative current
$\delta j$, however, causes an asymmetry in the domain lengths $d_-$,
$d_+$, given by
\be
\delta j \fr{\partial}{\partial j} \left. \fr{ d_+ - d_- }{b}
\right|_{j=0}  \approx
- \fr{\sqrt{2}}{\pi} \fr{E_0^2}{q^2}  \fr{\delta j}{u E_0} \sim
- \fr{ \delta  j}{u E_0} \exp(2d /b)   \:,
\ee
which in turn gives rise  to ascending or descending
potential profiles (Fig.~\ref{fig-formation}). The crucial point is
that for periods $\lambda = 
d_-+d_+$ larger than $b$ an exponentially small dissipative current $j
\sim uE_0 \exp(-\lambda / b)$ comes with a drastic change of the
potential-profil.

At first glance, these findings seem to give an explanation for residual
dissipative resistivity: an exponentially small dissipative current in $x$
direction with its necessarily ascending or descending potential
profile gives rise to a large Hall current (in the $y$-direction).  
However, this is not quite so simple: the dissipative current
of the stationary states flows from the lower to the higher potential, 
as can easily be verified within the mechanical
model. Therefore the residual resistance ascribed to these states
is negative. For the same reason these 
states are dynamically unstable. This can be checked by use of the
dynamical equation \Ref{dyn_equation}: A perturbation of
the domain sizes leads to currents that tend to increase the perturbation. 
It is not surprising that the transient currents involved in this
process are, again, exponentially small, $\propto
\exp(-\lambda/b)$. Note that the temporal evolution must therefore slow
down exponentially with increasing ratio $\lambda/ b$.   

Are these transient dissipative currents also responsible for the
observed residual resistance? To make this idea concrete, we analyze
in the following one scenario, in which we can explicitly relate the
transient currents to a transient residual resistance of the entire
system.  

\begin{figure}
\centerline{
\epsfig{file=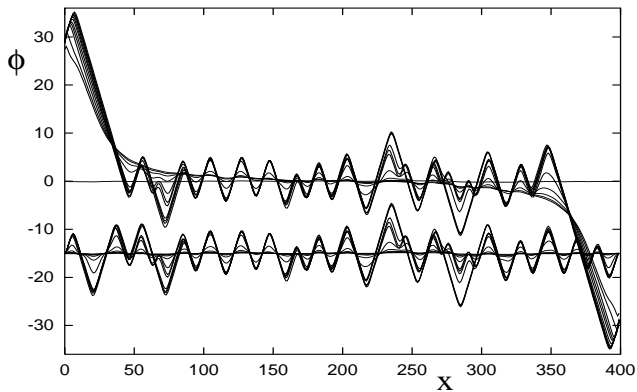,width=3.4in,height=2.1in}
}
%\vspace{.2in}
\caption{ Numercial results:
The electric potential developing from an initial random charge density (white
noise correlation) under microwave irradiation at finite (upper) and 
vanishing Hall voltage (lower curve, with a offset of -15 for better visibility).
The evolution has exponentially slowed down (time steps between
successive lines increase by a factor $\sqrt{2}$), however, peaks still
tend to merge (structure around $x\approx 250$). The screening length
of $s=65$ is larger than the period. In response to a finite Hall
voltage two large edge domains form, which move slowly inwards. 
Parameters in natural units ($l=1$, $\omega_c=1$, $e^2/l =1$):
$D=1.5$, $\delta t = 0.1$, $u = 0.5$, $E_0 = 1.0$  
}
\label{fig-num1}
\end{figure}

We consider a symmetric DC bias (Hall voltage) $\Delta \phi$. Knowing
that the stationary solutions 
discussed above (cf. Fig.~\ref{fig-formation}) are not physically
relevant,  it is not clear from the start how the potential drop
happens across the system. We have approached this problem by numerical
simulations of the system's dynamics, governed by
Eq.~\Ref{j_relation}, the equation of continuity, and Eq.~\Ref{1d_phi}
using the Green function of the Coulomb potential with different
screening lengths $s$. The lower curves in Fig.~\ref{fig-num1} are
the results for zero bias. The resemblance to the theoretical
profiles with zero current is obvious. Note that the profiles keep on
developing, albeit exponentially slowly with increasing $\lambda$.  
The upper curves in Fig.~\ref{fig-num1} are typical potential
formations for non-zero bias. 
By injecting resp.~extracting charges the electric potentials at the
system boundaries were fixed to given reference voltages.  
Significantly, in the bulk the profile looks the same as in the
case of vanishing voltage, and the entire potential drop happens near
the edges over two large edge domains. It can
also be seen that these domains slowly but steadily move inwards.  

It turns out that the transient profile with applied bias
in Fig.~\ref{fig-num1}, too, can  be analyzed in the mechanical
analogy, when the dynamical term in 
Eq.~\Ref{dyn_equation} is included as an $x$ dependent (anti-)friction term.
In mechanical terms one domain corresponds to a half cycle of
oscillation in the potential $V(E)$, with the domain length set by
the energy of the particle. To proceed we use a shock-wave ansatz
$E(x,t) = E(x-vt)$ \cite{volkov_kogan}, with the velocity $v$
piecewise constant in $x$. By means of this the time derivative of
Eq.~\Ref{dyn_equation} becomes a (anti-)friction term 
$ - \gamma \partial_x E(x-vt)$, with coefficient $\gamma = v/D$. 
This term extracts or supplies energy, and allows changes of
the domain length at the expense of a time dependence in the
corresponding potential profiles. 
By inspection of Fig.~\ref{fig-num1} one finds that significant
friction is needed at the edge domains only. In order to switch from a
large  edge domain length $r$ to the smaller bulk period $\lambda$ an
exponentially small  
energy of order $\exp-\lambda/b$ has to be removed. One arrives at
this result with the same ideas that led to Eq.~\Ref{exponential}.
Consequently, the edge domain must drift with a velocity $v \sim
\exp(-\lambda/b)$ into the system. The additional charges needed to
move the edge domain imply an exponentially small transient
current $I \propto r \exp(-\lambda/b)$. 
The domain on the right edge can be described similarly. Note that
for large $\Delta \phi$ there is Ohmic behavior in the sense
that the current is proportional to the Hall voltage, because  $r$ is
proportional to $\Delta \phi$. 

It is apriori not impossible that the observed resisidual resistance
is in fact transient, as in our example. However, this has not been
seen yet in experiment. If the dissipative current turns out to
be persistent, the 1D treatment presented here cannot provide an
answer, and a generalization to two dimensions might be necessary. 
Nevertheless, based on the preceding discussion we believe that the
exponent $\lambda/b$ generally governs the suppression of the
dissipative current, as $\lambda$ and $b$ are the only two fundamental 
lengthscales of the charge modulation. 
We expect this to be
independent of the particular choice of the Green function and
screening, provided the interaction is long ranged. This is confirmed
by numerics and further analytical considerations \cite{tobe}.  

We now relate $b$ to experimental parameters and discuss below
what limits $\lambda$. For the Coulomb interaction we find, by a simple
analysis using Eq.~\Ref{E_integral1D},  $b \sim  D/u $, which is a
microscopic 
length scale of the problem. The diffusion coefficient $D$ can be 
determined by Einstein's relation from the dark conductivity. 
Determining $u$, the dissipative conductivity in the dips, is more
difficult, because $u$ does not correspond to directly measured
quantities. However, an estimate can be obtained, if one assumes that the 
microwave enhancement and reduction of the conductivity is symmetric,
i.e that it is lowered in the dips by the same amount that it is
enhanced in the peaks (cf. \cite{durst}). Using the data
underlying Fig.~3 in Ref.~\cite{zudov}, we find $b \sim D/u$ 
of the order of the magnetic length or larger,  except for the
smallest resistances of the $1.1kG$ minima (see below).  
Furthermore, the resistivity enhancement (solid circles in Fig.~3 in
Ref.~\cite{zudov}) is linear in the inverse temperature, which
suggests the same for the reduced resistivities, hence $u \sim
C_0 + C_1/T$. 
Recall that by our hypothesis the residual resistance is
proportional to the factor $\exp(-\lambda/b) \sim \exp(-\lambda u /
D)$.  Assuming that $\lambda$ is a weak function of $T$, the $C_1/T$
dependence of $u$ gives rise to activated-like behaviour. 

It is clear that the magnetic length is a lower bound to $b$. When this
limit is reached at sufficiently low temperatues we expect a drastic
change in the $T$ dependence. Indeed, this limit seems to be reached in
the experiments; the estimated value for $D/u$ at the $1.1kG$ minimum
for the lowest temperatures in Fig.~3a in \cite{zudov} is about one
order of magnigtude smaller than the magnetic length. A corresponding
tendency of leveling off is visible in the data. 

Comparing the ``activation'' temperatures $T_0$ to $\lambda
C_1 /D$ shows that even moderate bulk periods $\lambda \gtrsim b$
can explain the large observed values for $T_0$. In addition, for
$\lambda$ and $D$ constant, the linear dependence of $C_1$ on the
magnetic field strength $B$, seen in the data \cite{zudov}, implies
$T_0 \propto B$. 

It remains an open question what determines the
domain width $\lambda$. A finite screening length $s$ is an obvious 
candidate. By incorporating screening into the mechanical
model we found evidence that $\lambda$ can assume certain fractions of
$4s$ which, however, can be arbitrarily small \cite{tobe}.
It is also possible that Landau level depletion, when a domain
becomes too large, significantly changes the local transport
properties to the degree that relation \Ref{j_relation} simply breaks down. 

A direct measurement of $\lambda$ would be a crucial test for the
domain picture. Furthermore, experimentally investigating 
correlations between $\lambda/b$ and the residual resistance allows a
test of our ideas.  

We thank R.~G.~Mani and M.~A.~Zudov for supplying us with experimental
information and data, and K.~von Klitzing and M.~R.~Zirnbauer for
stimulating discussions.


\begin{thebibliography}{h}
\bibitem{boundary_comment} For a system confined to $a \le x \le b$
  the integral extends from $a$ to $b$ and the boundary term
  $G(x-x')\rho(x')|_a^b\:$ has to be added.
\end{thebibliography}
\end{document}